# Evaluating Astronomy Literacy of the General Public

C. Love[1*], A. Murphy[2], S. Bonora[3]

[1]Department of Physics and Astronomy, Rowan University, Glassboro, NJ 08028
[2]Department of Neurosurgery, Thomas Jefferson University, Philadelphia, PA 19107
[3]Social Studies Department, Eastampton Community School, Eastampton, NJ 08060

## Abstract

A scientifically literate society is important for many different reasons, some of which include democratic and scientific topics. This study was performed in order to identify topics in astronomy and science in general that may not be well understood by the general public. Approximately 1,000 adults at a popular science museum in Philadelphia, PA completed True-False survey questions about basic astronomy concepts. The participants were also asked to provide their age, gender, and highest degree obtained. Although 93 ± 0.8% of the participants correctly answered that scientists can calculate the age of the Earth, only 58 ± 2% provided the correct response that scientists can calculate the age of the Universe. Some participants (30 ± 1%) responded that scientists have found life on Mars. Females scored an average total score of 78 ± 2%, whereas males scored an average 85 ± 1%. Participants with an age of 56 and over had an average score of 78 ± 4% compared to participants under the age of 56 that were found to have an average score of 82 ± 2%. Lastly, participants' highest degree obtained scaled with number of correct responses, with graduate level degree earners providing the largest amount of correct responses and an average score of 86 ± 2%.

## 1. Introduction

Many government and academic sources agree that a scientifically literate population is needed for a well-functioning, democratic, and technologically demanding society [1-11]. Knowledge of what scientists have accomplished and how science is done can help individuals understand problems in our society and make informed decisions when voting. As Laugksch and Spargo state, "Widespread scientific literacy of individuals is increasingly seen as being of vital importance for a number of different reasons-scientific, economic, ideological, intellectual, and aesthetic" [12]. Furthermore, a more informed public should make better decisions about research funding and even public safety concerns [13].

There is a concern in developed nations that there is a disparity between the scientific community and the general public [14]. Scientific information may get distorted, delayed, or even halted given the complex process of information transfer from scientists to the media/textbooks and finally to individuals of the general public. These individuals then must be able to understand and critically read about scientific concepts presented in both science textbooks and media articles [15-16].

The goal of this research is to determine the level of astronomy knowledge of the general public. If the scientific community understands what the public knows we can propose an effective strategy to build upon their current knowledge base. Previous studies have been done for assessing science literacy in educational atmospheres [8,15] as well as members of the general public [13-14,17-19]. To the authors' knowledge, no study has been performed using these specific topics in astronomy while attempting to probe the general public. This particular study focuses on modern questions that the public has been informed or misinformed of over the very recent past.

## 2. Methods

Large-scale surveys can provide baseline data and also, explore, describe, classify, and establish relations among variables [20]. For this study, approximately 1,000 adults at the Franklin Institute (a popular science museum) in Philadelphia, PA completed T/F survey questions about basic astronomy and scientific concepts. The participants were also asked to provide their age, gender, and highest degree obtained.

Our surveys were modeled after survey questions developed by Laugksch and Spargo based on Science for All Americans [4]. Furthermore, the surveys were created to be short and simple with only six true-false questions and ideally suited for inexpensive and easy administration [12]. The content of the survey questions were also designed to probe how well popular media has informed the public of more recent astronomical insights and discoveries that may not have been previously taught in the K-12 setting or college settings.

Question 1 on the survey stated "Scientists have discovered planets going around other stars," followed by Questions 2 and 3: "Scientists know pretty much all there is to know about the Universe" and "Scientists have found life on Mars." The last three questions (Questions 4-6) stated "Scientists can calculate the age of the Earth", "Scientists can calculate the age of the Universe, and finally "Science (peer-reviewed) journals are very similar to the NY Times or National Geographic." Participants were provided with a blank space to fill in their age and were asked to circle "Male or Female" and "Highest Degree Obtained: High School or Bachelor's or Master's or Doctorate or N/A." The complete survey can be found in Appendix A.

The procedure followed was to (1) ask the subject if they would be willing to fill out a short survey for a research project. If they agreed, they were (2) directed to the posted informed consent form. If they decided they would like to fill out the survey, we then (3) gave them a survey, pen, and clipboard, and they (4) completed the survey and (5) returned it when done to the best of their ability. No survey content related questions were answered while they were working on the questions and no collaborative answers were accepted.

The subjects were adults (over the age of 18) from the general public in attendance at the Franklin Institute. We attempted to attain a random sample of unique individuals for this research with this particular approach. However, in order for our subjects to access our booth and surveys they had to pay the admission fee for the Franklin Institute museum. We therefore, had a particular sample of the general public where according to a separate study [21] done by the museum's own internal polling, 58% of visitors to the Franklin Institute were very interested in science and 39% were somewhat interested (N=297). However, 83% of visitors were neither amateur nor professional scientists (N=288). Also, many visitors to the museum are on vacation or sightseeing through the city with 70% staying overnight in a hotel/motel and 23% staying at a friend's/relative's house (N=98). The cost of attendance to the museum is $16.50 for adults

with additional costs for traveling and specialty exhibits. The approximate annual household income was over $100,000 for 40% of visitors, whereas 31% had an approximate annual household income between $60,000 and $80,000, and finally 30% reported less than $60,000 approximate annual household income (N=232) [21].

## 3. Results and Discussion

As shown in Table 1, 48% of our participants were male and 52% were female (N=990). The museum's polling provided similar results with slightly higher percentage of visitors being female (53%), whereas 47% were reported as males (N=3973) [21].

According to the same museum internal polling (N=282), 31% of visitors received a graduate/professional degree compared to 33% of our participants earning either a master's or doctorate degree. Our results show that 42% of participants received a bachelor's degree and 24% received a high school diploma compared to the museum's internal polling that reported 40% and 30% respectively [21]. The results of for both of these studies are especially interesting when compared to a study on the general United States population. According to the National Center for Education Statistics, approximately 28% of adults over the age of 18 have earned a bachelor's degree or higher [22].

### Table 1: Numbers of Partipants in Each Category

| Total | Gender | | Age | | |
|---|---|---|---|---|---|
| 1047 | Male | Female | 18-35 | 36-55 | ≥55 |
| | 477 | 513 | 403 | 516 | 109 |
| **Highest Degree Obtained** | | | | | |
| N/A | HS | Associates | Bachelor's | Master's | Doctorate |
| 13 | 222 | 20 | 413 | 219 | 105 |

*Table 1: The number of participants in each gender, age, and degree groups. Note that the total reported in the top left corner is the total number of surveys received and may not have been completely filled out to include gender, age, and highest degree obtained.*

### Table 2: Percentages of age groups comparison

| Age Reported | 20-29 | 30-39 | 40-49 | 50-59 | 60+ |
|---|---|---|---|---|---|
| This Study (N=1005) | 22% | 30% | 30% | 10% | 8% |
| Museum Polling (N=949) | 20% | 23% | 32% | 8% | 17% |

*Table 2: The percentage of participants for each age group is reported alongside results from the museum's internal polling [21].*

Similarly, the percentages of age groups polled are shown in Table 2 and they are directly compared to the museums internal polling. Note that although our participants were over the age of 18, only those shown over the age of 20 are shown for comparison reasons. Generally, both polls show a larger population in the 30-49 region and a smaller 50 and older population.

All of the survey responses for this study are shown in Figure 1. At least 80% of the participants provided the correct response for four of the survey questions, with the question "Scientists know pretty much all there is to know about the Universe" being the most commonly correctly answered at 97 ± 0.5%. The two questions that participants most commonly answered incorrect (less than 70%) are two of the more controversial questions: "Scientists have found life on Mars" and "Scientists can calculate the age of the Universe".

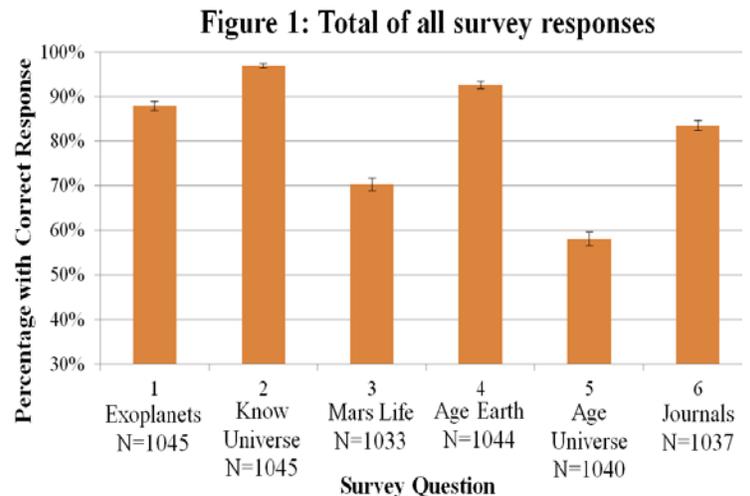

*Figure 1: A bar graph showing the percentage of all participants that provided the correct response. Each of the six survey questions are shown along with the number of participants. Note this number may vary due to incomplete surveys.*

Thirty percent (± 1%) of participants responded that scientists have found life on Mars. This particular question seemed to cause the most confusion and sparked the many discussions and arguments. Many participants wanted clarification of this question and discussed things like "evidence for": "bacteria", "single-celled organisms", "fossils", and "water". No one mentioned to us their belief in alien abductions or green Marian men and some people specifically said that is not what they meant. Some of the participants knew they heard something but where not sure what they heard and some were very adamant that life had been found on Mars.

Another interesting result from the survey responses showed that although 93 ± 0.8% of the participants correctly answered that scientists can calculate the age of the Earth, only 58 ± 2% provided the correct response that scientists can calculate the age of the Universe. No controversial comments came from discussing the age of the Earth and it was rarely even mentioned. However, some responses about the age of the Universe were as follows, "I know they think they can", "That's a trick question", "That's impossible", "The Universe is infinite", "Approximately", "We have no idea how big it is", "They can guess", and "That's philosophical". Further research should be done to explore some of the emotions expressed over this question.

The percentage correct for each question is separated by gender and shown in Figure 2. Females were found to score an average total score of 78 ± 2%, whereas males scored an average 85 ± 1%. Although not always outside of the error bars, males scored higher than females for each question.

In Figure 3, the percentage correct for each question is separated by age group. For each question, the scores between age groups are very similar and are within the error bars. However, participants with an age of 56 and over had an average score of 78 ± 4% compared to participants under the age of 56 that were found to have an average score of 82 ± 2%. The statistics need to be improved here and the number of participants older than 55 are roughly one-eight the number of participants 55 and younger.

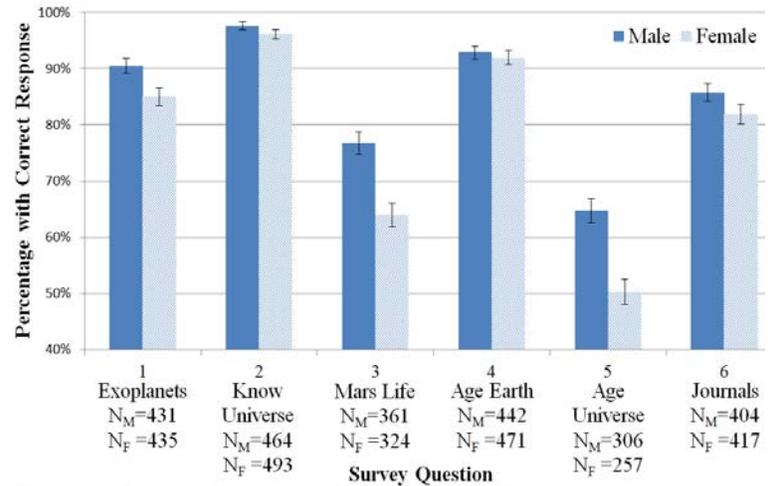

*Figure 2:* A bar graph showing the percentage of participants in three different gender groups that provided the correct response. Each of the six survey questions are shown along with the number of participants in each gender group. Note this number may vary due to incomplete surveys.

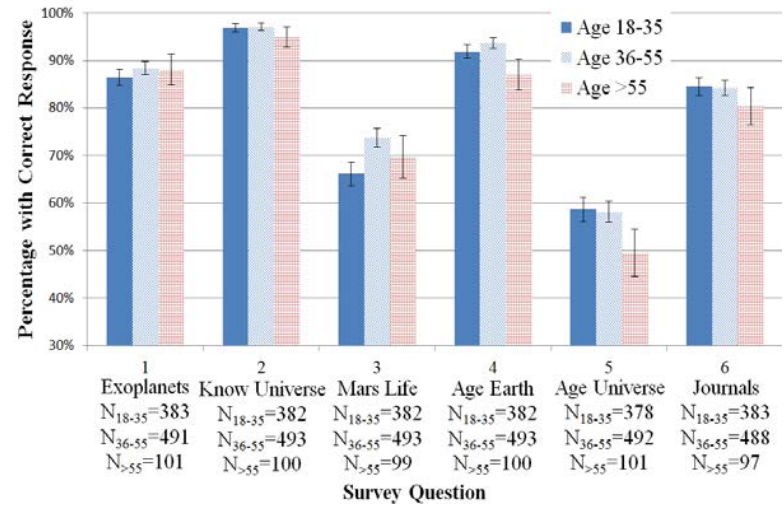

*Figure 3:* A bar graph showing the percentage of participants in three different age groups that provided the correct response. Each of the six survey questions are shown along with the number of participants in each age group. Note this number may vary due to incomplete surveys.

Highest degree obtained showed the most expected results. The higher the participant's degree, the higher number of correct responses provided. This is particularly apparent when looking at each question separately as shown in Figure 4. It should be noted that there was no category for associate degrees or some college. Occasionally participant's shared what type of higher degree they have obtained and some of their responses included MEd., MBA, JD, MD, and Doctor of Theology. Therefore, although these participants have higher degrees, they are not necessarily more scientifically educated than participants without higher education. This suggests that participant's with higher degrees are generally more knowledgeable and/or more informed than individuals that do not have higher degrees.

## 4. Conclusions

Generally, participants performed better than expected, both by the authors' and by the participants themselves. Some interesting and scientific discussions stemmed after participants completed the surveys. Most participants wanted to know the answers. Some participants claimed they were completely guessing but then they got most of the questions correct.

The error reported in this study is the standard error, taking into consideration the number of participants, which may have varied from subgroup to subgroup. It should be mentioned that the authors' believe there is even more error induced by several factors such as, international visitors where English is not their first language, collaboration and/or cheating, and guessing (both educated and completely random).

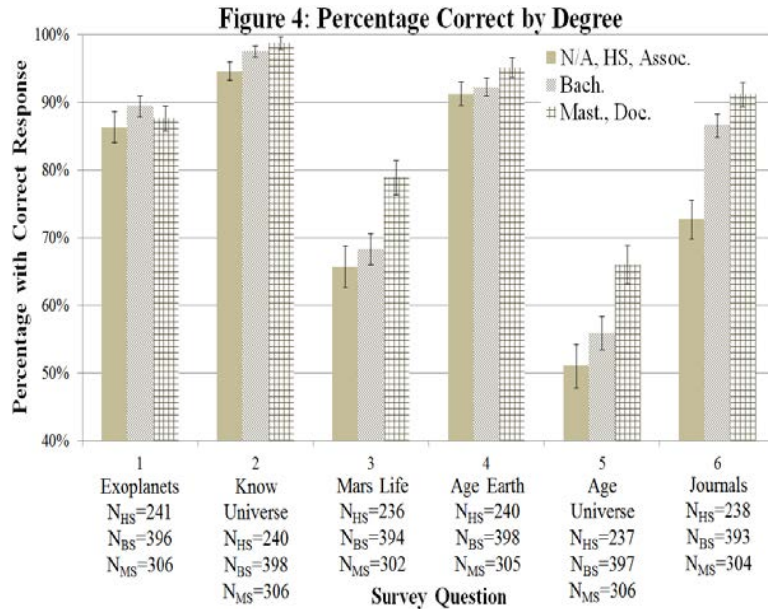

*Figure 4:* A bar graph showing the percentage of participants in three different degree groups that provided the correct response. Each of the six survey questions are shown along with the number of participants in each degree group. Note this number may vary due to incomplete surveys.

## 5. Future Work

More statistics would allow greater confidence in order to determine possible trends. More specifically, the relationship between age and percentage correct and also the relationship between highest degree obtained and percentage of correct responses were well within the errors reported. More surveys should be collected in the exact same manner to increase the statistics.

Another exciting option would be to use the exact same survey at a different location where individuals do not have as high of an approximate annual household income, where they are not interested in science, and where they do not have as many degree holders. Finally, the surveys should be edited in future studies to include individuals who have some college or other types of degrees.

## Acknowledgments

The authors would like to thank Liz Kelley, Mickey Maley, Jamie Collier, Minda Borun, and everyone at the Franklin Institute who helped make this project possible. We would also like to thank Jon Miller and Kim Bonora for their assistance in handing out surveys.

# Appendix A: Survey in complete form

*Instructions: Please fill in age and circle all other answers. Return when finished. Thank you!*

Age:________                    Male or Female

Highest Degree Obtained: High School or Bachelor's or Master's or Doctorate or N/A

T or F  Scientists have discovered planets going around other stars.

T or F  Scientists know pretty much all there is to know about the Universe.

T or F  Scientists have found life on Mars.

T or F  Scientists can calculate the age of the Earth.

T or F  Scientists can calculate the age of the Universe.

T or F  Science (peer-reviewed) journals are very similar to the NY Times or National Geographic.